\documentclass[prb,twocolumn,amsmath,amssymb]{revtex4}

\usepackage{graphicx,color}
\usepackage{multirow}

\begin{document}

\title{Low-temperature and high magnetic field dynamic scanning capacitance microscope}

\author{A. Baumgartner}
\email{andreas.baumgartner@unibas.ch}
\author{M.E. Suddards}
\author{C.J. Mellor}
\affiliation{
School of Physics and Astronomy, University of Nottingham, Nottingham NG7 2RD, UK\\}

\date{\today}

\begin{abstract}
We demonstrate a dynamic scanning capacitance microscope (DSCM) that operates at large bandwidths, cryogenic temperatures and high magnetic fields. The setup is based on a non-contact atomic force microscope (AFM) with a quartz tuning fork sensor with non-optical excitation and read-out for topography, force and dissipation measurements. The metallic AFM tip forms part of an rf resonator with a transmission characteristics modulated by the sample properties and the tip-sample capacitance. The tip motion gives rise to a modulation of the capacitance at the frequency of the AFM sensor and its harmonics, which can be recorded simultaneously with the AFM data. We use an intuitive model to describe and analyze the resonator transmission and show that for most experimental conditions it is proportional to the complex tip-sample conductance, which depends on both the tip-sample capacitance and the sample resistivity. We demonstrate the performance of the DSCM on metal disks buried under a polymer layer and we discuss images recorded on a two-dimensional electron gas in the quantum Hall effect regime, i.e. at cryogenic temperatures and high magnetic fields, where we directly image the formation of compressible stripes at the physical edge of the sample.
\end{abstract}

\maketitle

\section{INTRODUCTION}

For the investigation of individual electrically active nanostructures on surfaces the scanning tunneling microscope (STM) has become the standard tool.\cite{Binning_Rohrer_RevModPhys71_1999} However, its application is essentially limited to conducting samples on a surface. For layered semiconductors or capped nanostructures considerable efforts have been made to probe local properties by various other scanning probe methods at cryogenic temperatures and possibly in high magnetic fields. A prolific example is the investigation of the quantum Hall effect (QHE) in two-dimensional electron gases (2DEGs), where various scanning probe experiments have been demonstrated.\cite{Tessmer_Ashoori_Nature392_1998, Ahlswede_Weitz_PhysicaB298_2001, Ilani_Yacoby_Nature427_2004, Baumgartner_PRB76_2007}

An intuitive quantity to investigate is the capacitance between the fine metallic tip of an atomic force microscope (AFM) and the sample. Direct scanning capacitance microscopy (SCM) measurements in contact mode have been demonstrated for scanning,\cite{Gomila_Fumagalli_JAP104_2008, Fumagalli_APL91_2007} however it is a formidable task because the local capacitance is usually many orders of magnitude smaller than the cable and stray capacitances. Bridge measurements are possible,\cite{Smoliner_APL92_2008, Brezna_Smoliner_APL83_2003} but are rather slow and need readjusting for larger capacitance variations. The measurement of the complex impedance between tip and sample and simultaneous STM and SCM measurements have also been demonstrated.\cite{Lanyi_RSI73_2002, Lanyi_RSI65_1994} In AFM experiments, the voltage dependent part of the force between the tip and the sample can be interpreted in terms of capacitance.\cite{Kimura_APL90_2007, Kobayashi_APL81_2002} For samples with a tip-potential dependent capacitance it is possible to use a low-frequency ac voltage between tip and sample to modulate the capacitance in order to discriminate the local signal from the stray part. Dopant characterization in semiconductor structures\cite{Khajetoorians_JAP101_2007, Williams_AnnualRevMatSci_1999} is a well-established application of this technique.

In our experiments we use a radio frequency (rf) resonator in the form of a tuned filter coupled to the tip of an AFM for the capacitance detection.\cite{Williams_APL_1989, Matey_Blanc_JAP_1984} We combine this method with a mechanical modulation technique\cite{Bugg_King_JPhysE21_1988, Goto_Hane_RSI_1996, Naitou_APL85_2004} in an AFM that works at cryogenic temperatures and high magnetic fields. This combination allows us to use the AFM in the dynamic mode, which reduces tip wear and sample damage considerably, compared to most other techniques, where the tip needs to be scanned very close to or in contact with the sample surface. We use a phase-locked loop (PLL) to control the sensor position, so that we can separately record the topography, frequency shift (force gradient) and the dissipation of the AFM sensor. Simultaneously, the complex transmission of the tuned filter circuit coupled to the AFM tip is measured. This transmission is modulated by the sample properties as well as by the tip-sample capacitance. In contrast to the stray capacitance the latter is periodically modulated at the frequency of the sensor motion, which allows us to measure very small capacitance changes at high bandwidths.

In this paper we present the design of our dynamic scanning capacitance microscope (DSCM) (part II) and discuss the capacitance detection with the help of a quantitative lumped circuit harmonic oscillator model, where we introduce the tip-sample electrostatic interaction as a perturbation and show that the measured signals are in first order proportional to the complex tip-sample conductance (part III). This is different to scanning spreading resistance experiments,\cite{DeWolf_APL73_1998} where the dc resistance between an AFM tip (in hard contact) and the sample is measured. In part IV we analyze the modulation technique using intuitive generic capacitance vs. distance characteristics. We demonstrate DSCM imaging under various conditions in section V: we analyze scans on buried micrometer sized metal discs and on a two-dimensional electron gas (2DEG) in the quantum Hall effect (QHE) regime, i.e. at cryogenic temperatures and high magnetic fields.

\section{ATOMIC FORCE MICROSCOPE}

\begin{figure}[b]{
\centering
\includegraphics{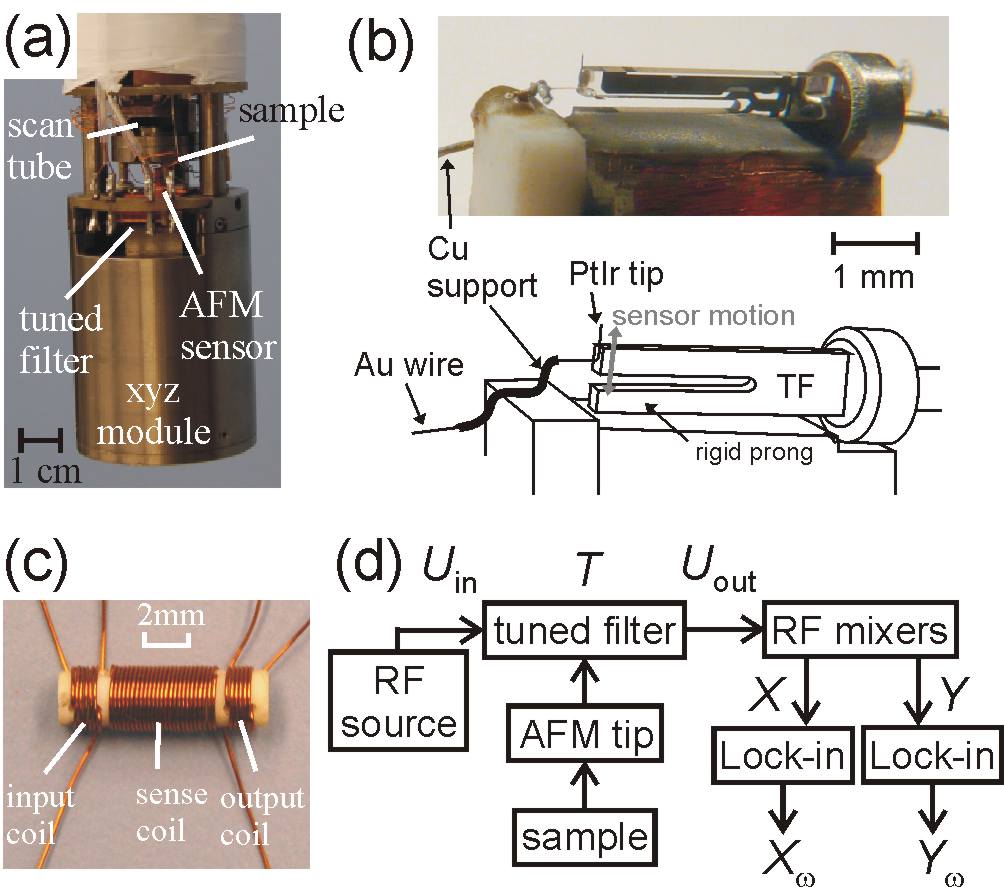}
}
\caption{(Color online) (a) Photograph of the AFM head. (b) Photograph and schematic of the TF sensor with PtIr tip and electrical contacts. (c) Cu coils forming the tuned filter. (d) Block diagram of the capacitance measurement.
\label{Figure1}}
\end{figure}

Our DSCM is based on a low-temperature AFM cooled in the variable temperature insert of a liquid helium cryostat, which allows continuous experimentation from $1.6\,$K to room temperature. A conceptually similar AFM design is discussed in Ref.~$22$. The dimensions of the copper AFM head can be seen in Fig.~\ref{Figure1}a. It is mounted on a $\sim 1.5\,$m long stainless steel tube, with $4$ coaxial cables and $10$ twisted pairs firmly attached to the probe and cooled continuously by the He flow. The coarse positioning is achieved by three commercial slip-stick motors and the scanning by a piezoelectric ceramic tube, with a lateral scan range that is reduced from $80$x$80\,\mu$m$^2$ at room temperature to $7$x$7\,\mu$m$^2$ at the base temperature of $1.6\,$K. The vertical range drops from $3\,\mu$m to $500\,$nm. The sample holder can be heated separately, which is used to keep the sample well above the microscope temperature before and during the cooldown to minimize condensation of water, nitrogen and oxygen on the surface. Various temperature sensors are installed for monitoring purposes. During the experiments the sample resides in the bore of a superconducting magnet, which generates static magnetic fields of up to $12\,$T perpendicular to the scan plane. We use home-built piezoelectric quartz tuning fork (TF) sensors with most of the magnetic parts removed and with an electrochemically sharpened $25\,\mu$m diameter PtIr wire attached to one prong. The other prong is rigidly glued to the AFM body, see Fig.~\ref{Figure1}b. A bridge circuit compensates the package capacitance. By modifying the tip-etch process we can influence the tip radius, which is useful because of the unavoidable compromise between lateral resolution and the capacitance sensitivity. The AFM sensor read-out and excitation are purely electrical, as required for the work with semiconductor samples at low temperatures. The excitation voltage is kept at the resonance frequency $f_{\text{TF}}\approx 32\,$kHz of the sensor by a phase-locked loop (PLL) and an additional feed-back loop keeps the mechanical amplitude of the sensor constant. This allows us to separately measure the frequency shift and the dissipation of the sensor.

For some of our experiments it is crucial to know the mechanical oscillation amplitude, $A$, of the AFM tip. We use a very simple technique to establish $A$, which works at any temperature and magnetic field: working at a fixed positive set-point for the frequency shift of the PLL an increase in $A$ is compensated by a reduced extension of the calibrated scan tube, $\Delta z$. The change in the tip position of closest approach to the surface can be neglected for the amplitudes considered here, because the force-distance curve is very steep at small distances from the surface. Plotting $\Delta z$ as a function of the TF current, $I_{\text{TF}}$, we find a linear relation for the range of $5\,\text{nm}<A<100\,$nm. For smaller amplitudes the PLL becomes unstable due to mechanical and electronic noise in the system. However, we expect the relation between $A$ and $I_{\text{TF}}$ to hold also in this regime. From such measurements we extract the piezoelectric coupling constant $\alpha=\frac{I_{\text{TF}}}{4\pi f_{\text{TF}}\cdot A}\approx5.4\,\mu$C/m, which is essentially the same at all temperatures and magnetic fields and comparable to values in the literature.\cite{Rychen_Ihn_Enssin_RSI71_2000}

\section{CAPACITANCE DETECTION}

In order to measure the tip-sample capacitance, $C_{\text{ts}}$, the metallic AFM tip is connected by a short and narrow gold wire (see Fig.~\ref{Figure1}b) to a tuned rf-filter consisting of three copper coils wound on a ceramic cylinder, see Fig.~\ref{Figure1}(c). A block diagram of the capacitance measurement is shown in Fig.~\ref{Figure1}d. We use an rf-generator and pre-characterized rf-power splitters and attenuators (not shown) to apply a voltage $U_{\text{in}}$ to the excitation coil. This leads to a voltage $U_{\text{out}}$ at the pick-up coil, which is amplified by an rf-amplifier (not shown) outside the cryostat. The central sense coil is connected to the AFM tip and forms an rf-resonator together with the cable and stray capacitances. The tip is coupled capacitively to the sample, which alters the transmission characteristics of the tuned filter depending on the tip-sample capacitance and the sample characteristics. With two rf-mixers of about $1\,$MHz bandwidth we recover the real and imaginary parts of $U_{\text{out}}$ with respect to $U_{\text{in}}$, $X=\text{Re}(U_{\text{out}})$ and $Y=\text{Im}(U_{\text{out}})$. Before an experiment we adjust the excitation frequency $f_{\text{rf}}$ of $U_{\text{in}}$ to the resonance of the rf-resonator and tune the phase to obtain a maximum $X$ and zero for $Y$. Typically $f_{\text{rf}}\approx130\,$MHz in our setup.

We use the AFM in tapping mode, which introduces a periodic variation of the tip-sample distance and therefore of $X$ and $Y$. The Fourier components of $X$ and $Y$, $X_{\omega}$ and $Y_{\omega}$ at the TF resonance frequency $\omega=2\pi f_{\text{TF}}$ are measured with standard lock-in amplifiers (see Fig.~\ref{Figure1}d), which allows signal recording with a bandwidth sufficient for scanning. In a similar way the second harmonics of the signals at twice the frequency, $X_{2\omega}$ and $Y_{2\omega}$ can be recorded.

\subsection{Unperturbed rf-oscillator}

\begin{figure}[b]{
\centering
\includegraphics{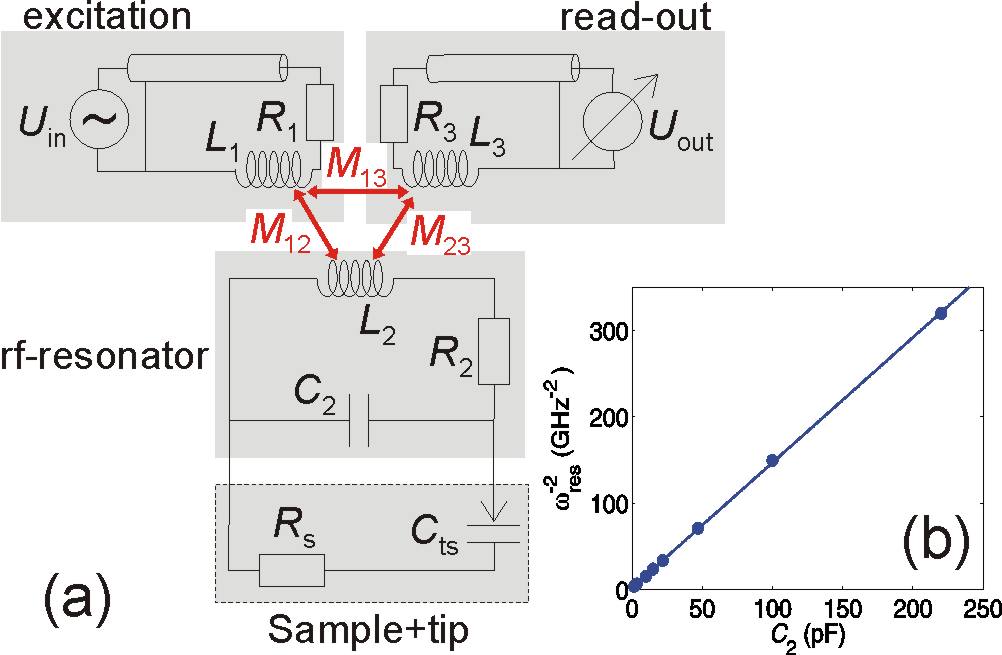}
}
\caption{(Color online) (a) Lumped circuit element representation of the rf-resonator coupled to the excitation and read-out circuit. The tip and the sample are coupled to the resonator in parallel to the stray capacitance $C_{2}$. (b) Inverse square of the resonance frequency as a function of calibrated capacitances.
\label{Figure2}}
\end{figure}

We now show that the intuitive model circuit in Fig.~\ref{Figure2}(a) reproduces the experimental rf resonance curve very well and that we can approximate the resulting transmission function $T=\frac{U_{\text{out}}}{U_{\text{in}}}$ by an LCR resonator. For this purpose we neglect the tip and the sample and re-introduce them later as a small perturbation. Because the coils, the tip and the relevant sample portions are small compared to the wavelength of the rf excitation signal, we model these parts of the circuit by lumped elements as shown in Fig.~\ref{Figure2}a. We include the coaxial cables as lossy wave guides.

The central coil, $L_{2}$, the self and stray capacitance $C_{2}$ of the coil and the AFM tip, and the wire resistance $R_{2}$ form an rf resonator, which is coupled to the excitation voltage $U_{\text{in}}\cos(\omega_{\text{rf}} t)$ by the inductor $L_{1}$ with resistance $R_{1}$ and a long coaxial cable. $L_{2}$ is also coupled to the detector coil $L_{3}$ with resistance $R_{3}$ and we measure the voltage via another long coaxial cable outside the cryostat. We are interested in the transmission $T$ at the resonance frequency of the unperturbed oscillator. The tip-sample system, represented in Fig.~\ref{Figure2}a by the sample resistance $R_{\text{s}}$ and the tip-sample capacitance $C_{\text{ts}}$ will be discussed in the next section.

\begin{figure}[t]{
\centering
\includegraphics{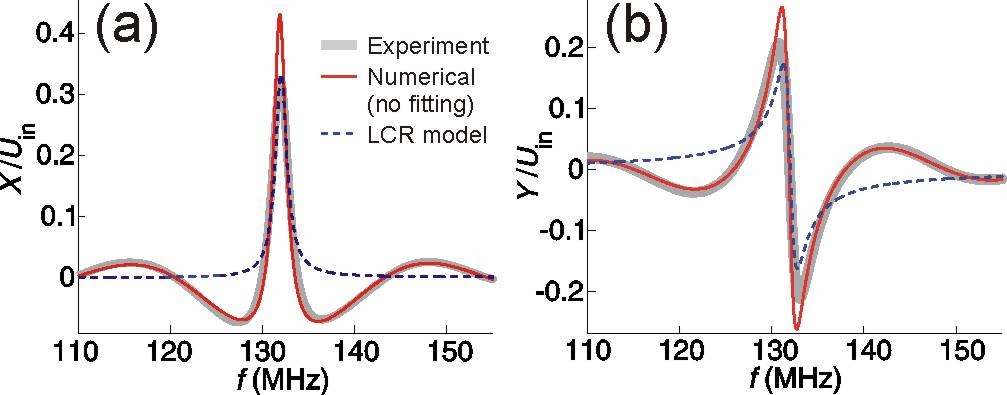}
}
\caption{(Color online) (a) Real ($X$) and (b) imaginary part ($Y$), normalized to $U_{\text{in}}$, obtained from the calculations in the text and from experiments.
\label{Figure3}}
\end{figure}

Using Kirchhoff's laws we find the following equation at any given frequency $f_{\text{rf}}=\omega_{\text{rf}}/2\pi$, relating the currents $I_{\text{j}}$ in each coil $L_{\text{j}}$ to $U_{\text{in}}$:
\begin{equation}
\left(
\begin{array}{c}
U_{\text{in}}\\0\\0
\end{array}
\right)
=
\left(
\begin{array}{ccc}
Z_{1} & -X_{12} & -X_{13} \\
-X_{12} & Z_{2} & -X_{23} \\
-X_{13} & -X_{23} & Z_{3}
\end{array}
\right)
\left(
\begin{array}{c}
I_{1}\\I_{2}\\I_{3}
\end{array}
\right)
\end{equation}
The diagonal elements are given by $Z_{1,3}=R_{1,3}+i\omega_{\text{rf}} L_{1,3}$ and $Z_{2}=R_{2}+i\omega_{\text{rf}} L_{2}+\frac{1}{i\omega_{\text{rf}} C_{2}}$, while the mutual inductances $M_{ij}$ give rise to the off-diagonal elements $X_{12}=i\omega_{\text{rf}} M_{12}$, $X_{23}=i\omega_{\text{rf}} M_{23}$ and $X_{13}=i\omega_{\text{rf}} M_{13}$. A comparison with experimental resonance curves shows that capacitive coupling can be neglected.
All the parameters in this matrix can be measured separately and the equation can be solved numerically. We incorporate the skin and proximity effects in the copper coils\cite{Savukov_JMR185_2007} and estimate the radiation losses to be negligible. The self- and mutual inductances can be determined by standard techniques.\cite{Avinor_Hasson_JoPE10_1977} To determine $C_{2}$ we soldered commercial high precision capacitors into the otherwise characterized circuit and measured the resonance frequency $\omega_{\text{res}}$ of the rf oscillator. In Fig.~\ref{Figure2}b $\omega_{\text{res}}^{-2}$ is plotted as a function of capacitance. $C_{2}$ in the lumped circuit model can be determined by interpolating the observed resonance frequency and we obtain $C_{2}\approx1.0\,$pF. From the calculated current in the coil $L_{1}$ we calculate the load impedance at the end of the first coaxial cable. Using standard formulas for transmission lines and the attenuation coefficient and the signal velocity, both given by the manufacturer, leads to the reflected power and to the voltage applied to the lumped element circuit. The measured output voltage is found as $U_{\text{out}}=Z_{0}\cdot I_{3}$, with $Z_{0}=50\,\Omega$ the impedance of the coaxial cable and the impedance matched mixers. The real and imaginary parts of $U_{\text{out}}$ are plotted as red lines labeled in Fig.~\ref{Figure3}. We note that no fitting parameters were used for this curve. However, the experimental curves in Fig.~\ref{Figure2} have a slightly lower quality factor than calculated. We attribute this to eddy current losses, e.g. in the cryostat parts. We incorporate these losses by using $R_{2}$ as a fit parameter, which reproduces exactly the experimental resonance curves in Fig.~\ref{Figure2} (not shown). This formalism also allows us to estimate the voltage on the AFM tip using $U_{\text{tip}}=\frac{1}{i\omega_{\text{rf}} C_{2}}I_{2}$. For a drive voltage of $U_{\text{in}}=1\,$mV we find $U_{\text{tip}}\approx24\,$mV at low temperatures. Such small values are desirable in low-temperature experiments on semiconductors because of possible persistent surface charging and have to be compared to tip voltages of the order of one volt necessary, for example, in depletion SCM experiments with similar sensitivity.\cite{Kobayashi_APL81_2002}

By varying the model parameters and comparing the resulting resonance curves with the experiment we find that the direct coupling between the excitation and pick-up coil is negligible in our setup. In the case of symmetric coupling ($M_{12}=M_{23}\equiv M$) and neglecting the effects of the coaxial cables one can find the following analytical solution
\begin{equation}
U_{\text{out}}=-\frac{\omega_{\text{rf}}^{2}M^{2}Z_{0}}{Z_{1}Z_{3}\left( Z_{2}+Z_{c}\right)} U_{\text{in}}
\label{analytic_response}
\end{equation}
with $Z_{c}=\frac{\omega_{\text{rf}}^{2}M^{2}}{Z_{1}}+\frac{\omega_{\text{rf}}^{2}M^{2}}{Z_{3}}$. A similar expression can be found for the tip voltage. The line shapes in Fig.~\ref{Figure3} and the linear dependence in Fig.~\ref{Figure2}b suggest that a harmonic oscillator model might be a good approximation. For $R_{1,3}<<\omega_{\text{rf}}L_{1,3}$, which is easily satisfied in our setup, we find the response of a damped harmonic oscillator in the form of a series LCR-resonator
\begin{equation}
U_{\text{out}}=\frac{W}{R_{\text{eff}}+i\omega_{\text{rf}} L_{\text{eff}}+\frac{1}{i\omega_{\text{rf}} C_{\text{eff}}}} \cdot U_{\text{in}},
\label{simplyfied_response}
\end{equation}
with $C_{\text{eff}}=C_{2}$, $L_{\text{eff}}= L_{2}-\frac{2M^2}{L}$, $R_{\text{eff}}=R_{2}+\frac{M^2}{L^2}(R_{1}+R_{3})$ and $W=-\frac{M^2}{L^2}Z_{0}$. The resulting resonance characteristics are plotted in Fig.~\ref{Figure3} as blue dashed lines. For a given $R_{2}$ no fitting is necessary to obtain these curves. Except for $R_{\text{eff}}$, the effective parameters change very little with temperature. Since $W$ is a constant at the resonance, it can not account for the phase change in the coaxial cables, hence the deviation from the experiment off-resonance in Fig.~\ref{Figure3}. Close to the resonance, however, the  LCR model reproduces all features relevant for our experiments and we restrict our further analysis to this model. Since we do not have access to the individual circuit elements at low temperatures, we use $R_{\text{eff}}$ as fit parameter to reproduce the rf-resonance curve. We also find an analytical expression for the maximum tip-sample voltage on resonance:
\begin{equation}
\left|U_{\text{tip}}\right|\approx \frac{M}{\omega_{\text{rf}} L_{1}C_{2}R_{\text{eff}}}\left|U_{\text{in}}\right|\approx \frac{M}{L_{1}}Q\left|U_{\text{in}}\right|,
\label{tip_voltage}
\end{equation}
where $Q$ is the quality factor of the resonator.

\subsection{Tip-sample interaction}

We model the effect of the tip and the sample on the rf resonator by introducing the tip-sample conductance $G_{\text{ts}}$ parallel to the stray capacitance $C_{2}$. The simplest representation is shown in Fig.~\ref{Figure2}, with a sample resistance $R_{\text{s}}$ and the tip-sample capacitance $C_{\text{ts}}$. We note that $G_{\text{ts}}$ may depend on external parameters like sensor position, magnetic field, or temperature.

Before we discuss the origin of $G_{\text{ts}}$ in more detail, we consider its effect on the output voltage of the tuned filter. Because one element of $G_{\text{ts}}$ is the tip-sample capacitance, $C_{\text{ts}}<<C_{2}$, the tip-sample interaction leads only to a very small perturbation of the rf resonator. To account for the tip and the sample we replace $(i\omega_{\text{rf}} C_{\text{eff}})^{-1}$ in Eq.~\ref{simplyfied_response} by $(i\omega_{\text{rf}} C_{2}+G_{\text{ts}})^{-1}$. By expanding the resulting equation into a Taylor series we find for $\omega_{\text{rf}}=\omega_{\text{res}}$ and $\left|G_{\text{ts}}\right|<<R_{\text{eff}}\omega_{\text{res}}^{2}C_{2}^{2}$
\begin{equation}
U_{\text{out}}\approx \frac{W}{R_{\text{eff}}}U_{\text{in}}\left[ 1-\frac{G_{\text{ts}}}{R_{\text{eff}}\omega_{\text{res}}^2C_{2}^{2}}\right]
\end{equation}
On resonance and far away from the sample surface we have $X=X_{0}\equiv\frac{W}{R_{\text{eff}}}U_{\text{in}}$, which leads to the following equation that relates the complex $G_{\text{ts}}$ to the variation in $X$ and $Y$ induced by the proximity of the sample:
\begin{equation}
G_{\text{ts}}\approx-K\cdot \Delta U_{\text{out}}\equiv-K\cdot (\Delta X + i\Delta Y)
\label{response2}
\end{equation}
with 
\begin{equation}
K=\frac{R_{\text{eff}}\omega_{\text{res}}^2C_{2}^{2}}{X_{0}}
\label{K}
\end{equation}
$K$ is characteristic for the setup and independent of the tip and sample characteristics. We note that $K$ also changes with temperature, mainly because of smaller material resistivities at low temperatures. By comparing Eqs.~\ref{response2} and \ref{K} with Eq.~\ref{tip_voltage} one finds that the signal level divided by the tip-voltage (sensitivity) can be improved, for example, by reducing the stray capacitance $C_{2}$.

$X$ and $Y$ both vary linearly with $G_{\text{ts}}$ and therefore allow us to measure the complex tip-sample conductance as a function of the tip position. Since the tip characteristics ideally do not change during a scan, variations are due to changes in the sample characteristics. We will discuss this point in more detail below. For well conducting samples $G_{\text{ts}}$ reduces to the purely imaginary conductance due to the tip-sample capacitance, $C_{\text{ts}}$. We find
\begin{equation}
C_{\text{ts}} \approx -\frac{K}{\omega_{\text{res}}} \cdot \Delta Y
\label{Rs_zero}
\end{equation}
for $C_{\text{ts}}<<R_{\text{eff}}\omega_{\text{res}}C_{2}^{2}$, much larger than typical changes of the tip-sample capacitance for reasonably sharp AFM tips. With the modulation technique described below we can measure voltages of $\sim 2\,$nV at a bandwidth useful for scanning, which results in a theoretical capacitance resolution of about $0.1\,$aF in our setup.

\subsection{Realistic tip-sample model}

It is quite intuitive that $U_{\text{out}}$ can be affected by the sample resistance. However, it is not obvious what the parameters $R_{\text{s}}$ and $C_{\text{ts}}$ in Fig.~\ref{Figure2} exactly mean. We therefore adapt a simple model\cite{Murray_VacTech_2007} to investigate the effects of the tip geometry and the resistivity and capacitance of the sample.

\begin{figure}[b]{
\centering
\includegraphics{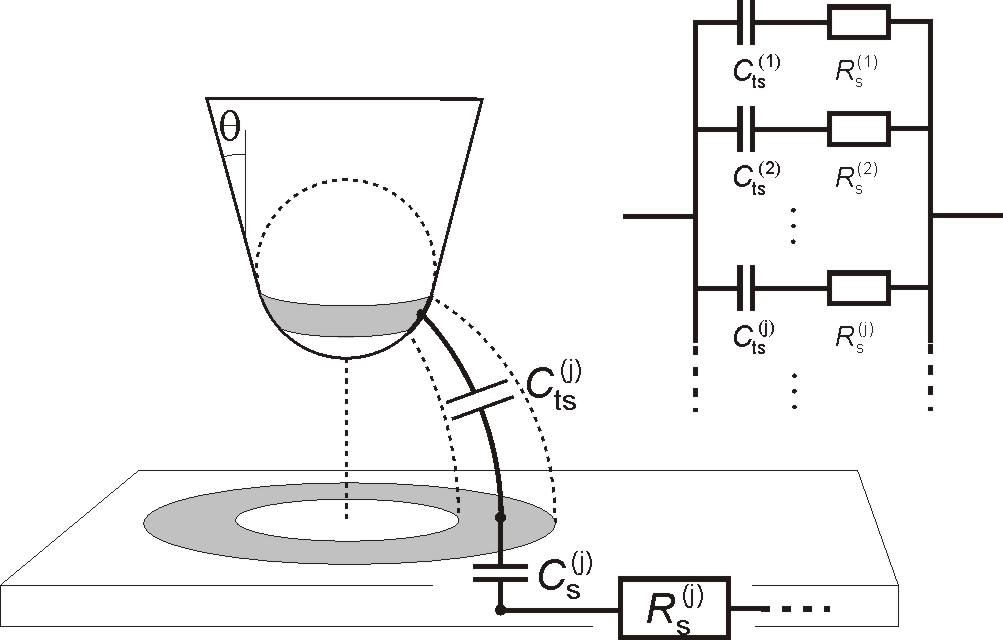}
}
\caption{Schematic capacitive coupling of a thin sample to an AFM tip modeled as a cone with a sphere at the end. Inset: tip-sample conductance as a series of parallel $RC$-elements.
\label{Figure4}}
\end{figure}

Fig.~\ref{Figure4} schematically shows the metallic tip modeled as a cone with an angle $\theta$ that ends in a sphere of radius $R$. We approximate the electric field lines as segments of circles orthogonal to the tip and the sample surface. The field lines define concentric areas labeled by $j$ on the tip and on the sample surface (shaded areas) which form a plate capacitor, $C_{\text{ts}}^{(j)}$, with a separation of the plates that corresponds to the arc length of the field lines.\cite{Murray_VacTech_2007} This is only a good approximation if the sample is metallic or has a large dielectric constant (e.g. GaAs) and if the tip-sample separation is not too large. For each area $j$ on the sample we can include a sample capacitance $C_{\text{s}}^{(j)}$ in series with $C_{\text{ts}}^{(j)}$, by which it is possible to include, for example, the capacitance of a dielectric layer or the quantum capacitance of the conducting layer, as might be relevant for experiments with semiconductor samples with a low density of states at the Fermi energy. For the present purpose we neglect the stray capacitance between the sample and the metallic sample holder.

Each of the composite capacitor elements is connected to ground by a possibly rather complicated current path, represented in Fig.~\ref{Figure4} by the resistor $R_{\text{s}}^{(j)}$. These lumped circuit elements can be thought of as a network of parallel RC elements (see inset of Fig.~\ref{Figure4}) which can be analyzed numerically. Here we only discuss the maximum (homogeneous) sample resistivity in a metallic, two-dimensional homogeneous plane,\cite{footnote1} for which we can use Eq.~\ref{Rs_zero}, i.e. for which we can neglect the sample resistivity and directly measure the tip-sample capacitance. We assume the sample to be circular with a well-defined local resistivity $\rho$ and to be grounded at the circumference at the distance $a$. For this case one finds the corresponding resistance from the outer electrode to the ring $j$ of radius $r_{j}$ as $R_{\text{s}}^{(j)}=\frac{\rho}{2\pi} \ln \left( \frac{a}{r_{j}} \right)$

Equation \ref{response2} reduces to Eq.~\ref{Rs_zero} if $\omega_{\text{res}}R_{\text{s}j}C_{\text{ts},j}<<1,\,\forall j$. Both, the $C_{\text{ts},j}$ and $R_{\text{s},j}$ decrease with increasing $r_{j}$, so that if the constraint holds for the innermost element and for the position of closest approach, it holds for all elements and positions. This first element we can approximate by a plate capacitor with appropriate radius and tip-sample separation (see Eq.~\ref{tip_radius} below). We estimate an upper limit of $\rho_{\text{crit}}^{(2D)}\approx 10^{7}\,\Omega_{\square}$. We obtain similar values for the analytical model with various tip radii and cone angles. For smaller resistivities we can use Eq.~\ref{Rs_zero}, e.g. for thin gold layers or on graphite, as used below.

\section{CAPACITANCE MODULATION TECHNIQUE}

\subsection{Modulation mechanism}

\begin{figure}[b]{
\centering
\includegraphics{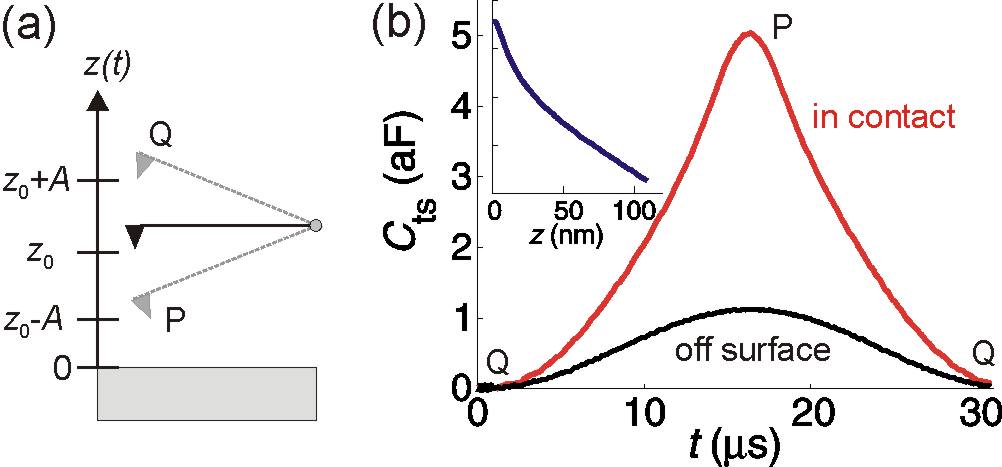}
}
\caption{(Color online) (a) Schematic tip motion and coordinate system. (b) $C_{\text{ts}}(t)$ trace averaged over $\sim 160,000$ cycles on a freshly cleaved HOPG surface. The red curve shows the capacitance when the AFM tip is `in contact', the blue when the tip is withdrawn $300\,$nm from the surface (`off surface'). The inset shows the $C(z)$ curve extracted from the data with the tip in contact. ($T=2.5\,$K, $U_{\text{tip}}\approx300\,$mV, $A\approx60\,$nm)
\label{Figure5}}
\end{figure}

We use the AFM in the `tapping' mode, i.e. the AFM tip oscillates with an amplitude $A$ and with a positive setpoint of the PLL ($\sim$ average force per TF cycle). This allows us to simultaneously record topographic and capacitance images while scanning with minimum tip wear. The mechanical oscillation modulates the tip position above the sample surface, $z(t)$ (see Fig.~\ref{Figure5}a), which in turn leads to a modulation of the tip-sample conductance $G_{\text{ts}}(t)$, introduced in the previous section, at the frequency of the TF resonance and its harmonics. Figure 5b shows two direct measurements of $C_{\text{ts}}(t)=\frac{K}{\omega_{\text{res}}}Y(t)$ on a freshly cleaved highly ordered pyrolytic graphite (HOPG) surface at $T=2.5\,$K as a function of time, $t$, averaged over $\sim160,000$ AFM oscillations with amplitude $A\approx 60\,$nm. `Off surface', i.e. when the tip is withdrawn by about $300\,$nm from the surface, $C_{\text{ts}}(t)$ exhibits a cosine form with a relatively small amplitude.  When the AFM sensor is `in contact' (controlling), the signal amplitude increases strongly and the curve deviates from the cosine shape. The maximum tip-sample capacitance occurs when the tip is closest to the surface (position P in Fig.~\ref{Figure5}a) and the minimum when the tip is furthest away (position Q). From such measurements we can directly reconstruct an approximate $C(z)$ curve, as is shown in the inset of Fig.~\ref{Figure5}b for the data `in contact'. However, this method of obtaining a $C(z)$ characteristics is extremely slow and we will discuss a more convenient technique elsewhere.

Using lock-in amplifiers we demodulate the outputs of the mixer stage, $X(t)$ and $Y(t)$ (see Fig.~\ref{Figure1}d) and record their respective modulation amplitude, $X_{\omega}$ and $Y_{\omega}$, at the tuning fork frequency, $\omega$. We notice that it is at this stage that for each position in a scan the large, time-independent background capacitance is removed. $X_{\omega}$ and $Y_{\omega}$ are equivalent to the Fourier-cosine coefficient of the signal and are related to the Fourier coefficient of the tip-sample conductance modulation $G_{\text{ts},\omega}$ by Eq.~\ref{response2}. For small variations the latter is related to the capacitance modulation due to the tuning fork motion, $C_{\omega}$, by
\begin{equation}
G_{\text{ts},\omega}\approx \frac{dG_{\text{ts}}}{dC_{\text{ts}}}\cdot C_{\omega},
\label{Gts_from_Cw}
\end{equation}
which requires a model for $G_{\text{ts}}$. A similar equation holds for higher harmonics of $C(t)$ if $C(z)$ is non-linear. For well conducting samples Eq.~\ref{Gts_from_Cw} reduces to Eq.~\ref{Rs_zero}, which leaves us with investigating $C_{\omega}$.

In the remaining part of this section we will consider the effect of a $C(z)$ dependence of the form 
\begin{equation}
C_{\text{ts}}(z)=C_{0}-B_{1}\cdot z + \frac{B_{2}}{z},
\label{C_vs_z}
\end{equation}
where $C_{0}$ accounts for large, but constant stray capacitances and can be ignored for our purposes. $-B_{1}z$ is a linear approximation to the large, but slowly varying capacitance of the bulk of the sensor. The last term accounts for the diverging part of $C(z)$ when the tip is very close to the surface. In this case the signal is dominated by the foremost part of the tip, which approximately forms a plate capacitor with the corresponding sample area. The $B_{2}$-term therefore has a local origin on the scale of the tip radius. We can define an effective area of this plate capacitor, $A_{\text{eff}}$, by $B_{2}\equiv\epsilon\epsilon_{0}A_{\text{eff}}$ from which we can estimate the effective tip radius
\begin{equation} R_{\text{tip}}=\sqrt{\frac{B_{2}}{\pi\epsilon\epsilon_{0}}}
\label{tip_radius}
\end{equation}
with $\epsilon$ and $\epsilon_{0}$ the relative and the vacuum permitivities. A typical value of $R_{\text{tip}}\approx40\,$nm can be extracted from the $C(z)$ curve in the inset of Fig.~\ref{Figure5} by fitting expression Eq.~\ref{C_vs_z}, with an offset of a few nanometers on $z$ as additional fit parameter, which we attribute to native oxide layers on the sample and the tip.

Because our AFM sensors are based on very stiff tuning fork resonators ($k\sim10^{4}\,\frac{N}{m}$) the tip-sample separation can be approximated by
\begin{equation}
z(t)=z_{0}+A\cdot \cos (\omega t),
\label{tip_motion}
\end{equation}
even when the AFM is controlling. Here we neglect possible periodic deformations of the tip and the sample surface.
First, we consider mechanical amplitudes small enough so that we can approximate $C_{\text{ts}}(t)\approx C_{\text{ts},0}(z_{0})+C_{\text{ts}}^{\prime}(z_{0})\Delta z(t) + \frac{1}{2}C_{\text{ts}}^{\prime\prime}(z_{0})\Delta z^{2}(t)$ over a complete sensor oscillation at a given average position $z_{0}$. The primes denote the spatial derivatives along $z$. For an arbitrary $C_{\text{ts}}(z)$ curve we find the first and second Fourier-cosine coefficients:
\begin{eqnarray}
C_{\omega}(z_{0})&=& AC_{\text{ts}}^{\prime}(z_{0})\\
C_{2\omega}(z_{0})&=& \frac{1}{4}A^{2}C_{\text{ts}}^{\prime\prime}(z_{0})
\end{eqnarray}
Because of the sharp increase of the tip-sample interaction force very near the surface we can assume that for a given PLL setpoint the distance of closest approach is essentially constant for different amplitudes. Using Eq.~\ref{C_vs_z} we find that these expressions are suitable only for  $A<<z_{0}/4$, i.e. rather far away from the surface, or not for the mechanical amplitudes used in our experiments. Therefore and for future reference we now consider larger amplitudes in the quite general case of Eq.~\ref{C_vs_z}. By substituting the tip motion of Eq.~\ref{tip_motion} in Eq.~\ref{C_vs_z} we find for $A<z_{0}$ the following series expansions for the even and odd Fourier-cosine coefficients (the first factor in the sums represents binomial coefficients):
\begin{equation}
C_{2m\omega} =
\frac{2B_{2}}{z_{0}}\sum_{k=m}^{\infty}
\left( 
\begin{array}{c}
	2k\\ k-m
\end{array}
\right)
\left( \frac{A}{2z_{0}} \right)^{2k}
\end{equation}

\begin{equation}
C_{(2m+1)\omega} =
-B_{1}A\delta_{m,0}
-\frac{2B_{2}}{z_{0}}\sum_{k=m}^{\infty}
\left( 
\begin{array}{c}
	2k+1\\ k-m
\end{array}
\right)
\left( \frac{A}{2z_{0}} \right)^{2k+1}
\end{equation}
For the first two coefficients we can find the more compact expressions
\begin{eqnarray}
C_{\omega} &=& -B_{1}A - \frac{2B_{2}}{A}\frac{1-\kappa}{\kappa} \label{Cw}\\
C_{2\omega} &=& \frac{2B_{2}}{z_{0}\kappa} \frac{1-\kappa}{1+\kappa}
\label{C2w}
\end{eqnarray}
with $\kappa=\sqrt{1-\frac{A^{2}}{z_{0}^{2}}}$. Because $z_{0}=A+d$ with $d\geq0$ the distance of closest approax, $0<\kappa<1$ has to hold. Eqs.~\ref{Cw} and \ref{C2w} show that $C_{\omega}$ depends on both, $B_{1}$ and $B_{2}$ and is always negative. For typical parameters ($\frac{B_{1}}{B_{2}z_{0}}>>4$) we find that the two terms are equal for $d\approx 0.04 A$. In contrast to $C_{\omega}$, $C_{2\omega}$ depends only on the short-range, non-linear part of $C(z)$ and is always positive. However, in the experiment we record only the absolute value of $X$ and $Y$ and we can not infer the sign from the data.

For $\kappa\approx 1$, i.e. $A<<z_{0}$, we find that $C_{\omega}$ measures $C'(z_{0})$ and $C_{2\omega}$ measures $C''(z_{0})$, as expected from above. However, for finite $\kappa$, e.g. when controlling over a metal surface, these simple relations are not correct anymore. Qualitatively, in many experiments the $C_{\omega}$ is on the same order as the capacitance variation between the positions of closest approach and the position furthest from the sample.

\subsection{Spatial contrast and resolution}

Only $C_{\text{ts}}(t)$ is time dependent due to the tip motion and generates non-zero Fourier components.\cite{footnote2} The TF amplitude is held constant during a scan and therefore does not generate features in the recorded images. The previous discussion suggests two fundamental mechanisms that can cause spatial contrast in $X_{\omega}$ and $Y_{\omega}$: 

\begin{enumerate}
	\item The $C(z)$ curve can be changed if the sample characteristics change as function of the lateral tip position. In Eqs.~\ref{Cw} and \ref{C2w} this would correspond to a change in the parameters $B_{1}$ and $B_{2}$. An important application of this mechanism could be the investigation of doping profiles in (capped) semiconductor structures. Two special cases will be important later:
	
	1a) A variation of the sample capacitance $C_{\text{s}}$ (see Fig.~\ref{Figure4}), e.g. due to a lateral variation of the density of states at the Fermi energy of a semiconductor or of the dielectric constant of a material.
	
	1b) In the case where the resistance from the tip position to ground is significant, a variation of the local resistivity also leads to a contrast in our experiments. Examples are certain regions of a 2DEG in the QHE regime.
	
	\item By changing the average vertical position of the sensor, $z_{0}$, the signals are generated by a different part of the otherwise unchanged $C(z)$ curve. Important examples are topographic features in a dielectric top layer, or the investigation of $C(z)$ curves.
\end{enumerate}

These mechanisms are not necessarily independent and have to be discussed for each experiment separately. For the lateral and the vertical resolution of the DSCM experiments we expect better values for the measurements of the higher harmonics of $Y$ because smaller parts around the tip apex are relevant. For example the $B_{1}$ and $B_{2}$ terms in Eq.~\ref{C_vs_z} stem from the bulk and the tip apex of the sensor, respectively, and we expect a higher lateral resolution for $Y_{2\omega}$ than for $Y_{\omega}$.

While in topographic imaging a small, possibly non-conducting protrusion at the end of the tip can enhance the spatial resolution considerably, in SCM experiments the total shape of the conducting tip has to be considered because of the long-range nature of the Coulomb interaction. Because of the high resolution of the capacitance measurement, the change in $Y$ due to the change in the relative tip-sample distance on a small topographic feature can be resolved, but does {\it not} reflect a lateral change in the electronic sample properties. These features can be resolved with the same lateral resolution as the topographic resolution. This `vertical' resolution has to be distinguished from the `real' lateral resolution: by comparing the simultaneously recorded topography image with the SCM data, the two origins of SCM features can easily be distinguished.\cite{Suddards_Baumgartner_PhysicaE40_2008} We expect a lateral resolution on the order of the tip radius given in Eq.~\ref{tip_radius} for tip-sample distances where the non-linear part of $C(z)$ is relevant. DSCM therefore combines a high resolution similar to contact SCM and a high reproducibility due to reduced tip and sample damage.

\section{DSCM EXPERIMENTS}

\subsection{Metal covered by dielectric}

\begin{figure}[t]{
\centering
\includegraphics{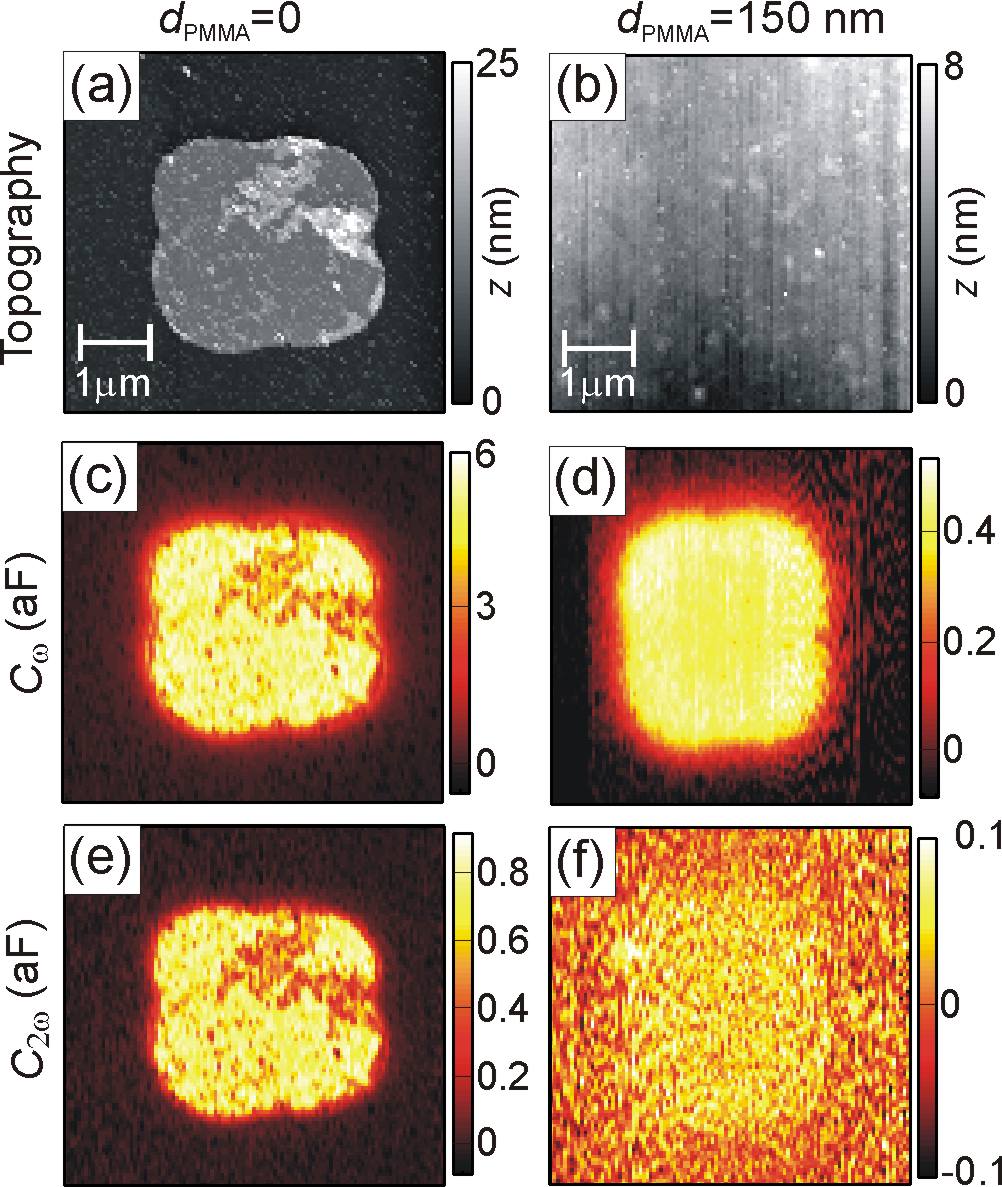}
}
\caption{(Color online) Dynamic SCM imaging on $\sim 20\,$nm thick and  $\sim 3\,\mu$m wide gold squares. Left column: bare sample, right column: similar sample with $\sim 150\,$nm PMMA deposited on top. The top row shows the respective topographic images, while the center and bottom rows show $C_{\omega}$ and $C_{2\omega}$ as a function of the lateral tip position. ($T=250\,$K, $\left|U_{\text{tip}}\right|\approx240\,$mV, $A\approx30\,$nm, data acquisition time per column:~$\sim7$ minutes)
\label{Figure6}}
\end{figure}

We performed DSCM experiments on gold squares of roughly $3\,\mu$m base length and $20\,$nm thickness, which were defined on a GaAs wafer by conventional photolithography and metal evaporation. The topographic AFM image of such a region taken at $T=250\,$K is shown in Fig.~\ref{Figure6}a. A similar sample was capped with a layer of polymethylmethacrylate (PMMA) of thickness $d_{\text{PMMA}}\approx 150\,$nm. The relatively flat surface of this layer (see Fig.~\ref{Figure6}b) was obtained by depositing a much thicker layer of PMMA and subsequent plasma etching. We note that from the topography image of this sample the presence of the Au square below cannot be inferred. The bandwidth of the measurement is detemined by the time constant of the demodulation lock-ins and was chosen $\sim30\,$Hz.

Simultaneously with each topography image we record $X_{\omega}$, $Y_{\omega}$ and $Y_{2\omega}$. In spite of the fact that the Au squares are not grounded, we can use Eq.~\ref{Rs_zero} to obtain the capacitance signals $C_{\omega}$ (Figs. c and d) and $C_{2\omega}$ (Figs. e and f), because the gold squares have a large self-capacitance and because gold is a very good conductor. No changes in $X_{\omega}$ could be resolved in these experiments.

On the sample without a PMMA layer, $C_{\omega}$ (Fig.~\ref{Figure6}c) increases by about $6\,$aF when the tip is moved from the unpatterned surface to the gold area, while the average tip-surface distance is kept constant. This signal difference allows us to image the gold area with a strong contrast. We note that the additional structures found in the topography image in the top right of the gold square leads to a {\it de}crease in $C_{\omega}$ with a similar lateral resolution as in the topography image.
In the image of $C_{2\omega}$ in Fig.~\ref{Figure6}e these topographic details lead to an even sharper contrast and also the gold square appears better defined. However, the maximum signal level in this case is only $0.8\,$aF.

We performed similar experiments on the sample covered by $150\,$nm of PMMA, where the topography image does not show the gold square beneath. However, the latter is clearly visible in the corresponding $C_{\omega}$ image in Fig.~\ref{Figure6}d. The signal level is reduced by a factor of about ten compared to the scan on the bare sample (Fig.~\ref{Figure6}c) and the lateral resolution is reduced. On this sample, the gold square can barely be resolved in the $C_{2\omega}$ image in Fig.~\ref{Figure6}f.

Away from the evaporated gold the nearest metallic surface is very far away and the capacitance variation over a sensor cycle is very small. On a metallic surface, however, the AFM tip motion takes place on a much steeper part of the $C(z)$ characteristics and the signal is large. The effect of the PMMA layer on the second sample is two-fold: the average tip-sample distance is kept at an additional $150\,$nm, which explains most of the signal change in $C_{\omega}$ (see Eq.~\ref{Cw}). However, to understand the changes in $C_{2\omega}$ the dielectric response of the PMMA layer has to be considered ($\epsilon\approx3$ at $130\,$MHz).

\subsection{2DEG in QHE regime}

Our microscope is designed to work at cryogenic temperatures and high magnetic fields. Here we present DSCM experiments performed at $T=2.4\,$K and at high magnetic fields, i.e. in the QHE regime of a 2DEG. The sample is a GaAs/AlAs heterostructure buried $\sim 50\,$nm below the surface with the donor atoms contained in a narrow quantum well $\sim39\,$nm below the surface.\cite{Shashkin_Kent_Eaves_PRB49_1994} The electron density extracted from the low-field Hall resistance is $n=4.0\times10^{15}\,\text{m}^{-2}$ and the zero-field mobility $\mu=10\,\frac{\text{m}^{2}}{\text{Vs}}$. A $100\,\mu$m wide Hall bar was defined by standard photolithography and wet chemical etching. The etched structure is $80\,$nm deep. We roughly compensate the contact potential difference between tip and sample by keeping the sample at a constant potential of $-0.3\,$V relative to the tip and we used the relatively small tip voltage $\left|U_{\text{tip}}\right|\approx35\,$mV to minimize local charging effects.
\begin{figure}[b]{
\centering
\includegraphics{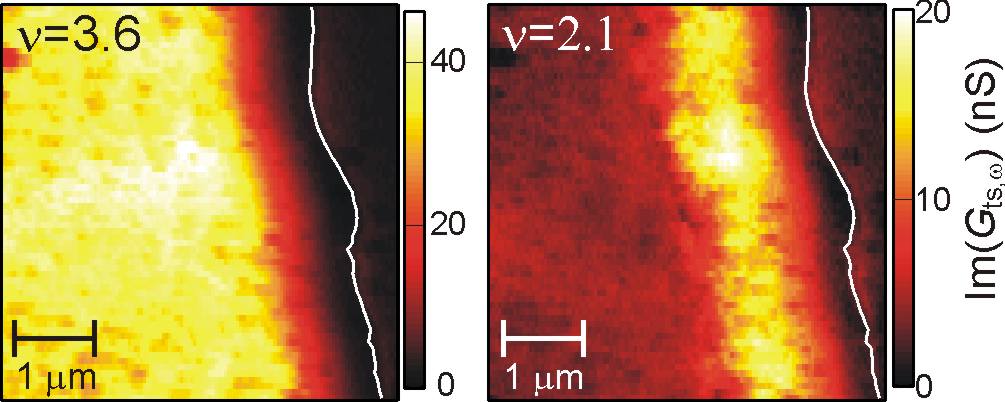}
}
\caption{(Color online) The imaginary part of the tip-sample conductance, Im$(G_{\text{ts},\omega})$, as a function of position at the edge of a Hall bar (the 2DEG is to the left of the white line). (a) bulk filling factor $\nu\approx3.6$ (b) $\nu\approx2.1$. ($T=2.5\,$K, $\left|U_{\text{tip}}\right|\approx35\,$mV, $A\approx30\,$nm)
\label{Figure7}}
\end{figure}
Figure 7 shows Im$(G_{\text{ts},\omega})$ derived from $Y_{\omega}$ for two scans on the same area taken at magnetic fields of $B=4.6\,$T and $B=7.9\,$T, which correspond to the bulk Landau level filling factors $\nu\approx 3.6$ and $\nu\approx 2.1$ of the 2DEG, respectively. The edge of the Hall bar is indicated by a white line extracted from the topography image. The 2DEG resides to the left. At non-integer filling factors, e.g. at $\nu=3.6$ shown in Fig.~\ref{Figure7}, we find that Im$(G_{\text{ts},\omega})$ is homogeneous as long as the tip resides over the 2DEG and drops strongly outside the Hall bar. The signal variations on length scales smaller than hundred nanometers can be traced back to topographic features as discussed before\cite{Suddards_Baumgartner_PhysicaE40_2008} and will be ignored for the rest of the paper. A homogeneous signal of similar strength can be observed at zero magnetic fields. Near integer filling factors, e.g. at $\nu=2.1$ in Fig.~\ref{Figure7}, Im$(G_{\text{ts},\omega})$ becomes inhomogeneous and the image shows a stripe of increased signal strength along the sample edge, while the signal drops to the level of bare GaAs (off the 2DEG) when the tip is moved onto the bulk of the 2DEG. We identify the stripe as related to the compressible regions that develop at the sample edges due to the formation of Landau levels in the 2DEG.\cite{Chklovskii_Shklovskii_Glazman_PRB46_1992} We will present more experiments and discussions on this topic elsewhere. Here, these experiments serve as an example for a sample where the average distance from the tip to the conducting plane is constant, but a magnetic field induces changes in the local density of states at the Fermi energy and the local resistance to ground. It is not a priori clear what causes the signal contrast in the presented images. A more involved analysis shows that in the areas with large $G_{\text{ts},\omega}$ the resistive part can be ignored and the signal strength is determined by the quantum capacitance of the 2DEG. This, however, is not the case, at the edge of the stripe. We will discuss such experiments elsewhere in more details, e.g. the imaging of the innermost incompressible stripe in $X_{\omega}$ and the effect of the rf power on the observed structures.

\section{CONCLUSION}

We have presented a dynamic scanning capacitance microscope based on the modulation of the transmission characteristics of a tuned filter by the mechanical motion of the tip of a non-optical AFM working in dynamic mode. We demonstrate an intuitive and, where it is useful, quantitative analysis of the resonator and show that the measured quantities can be interpreted as the Fourier-cosine coefficients of the periodically modulated complex tip-sample conductance. The modulation allows us to separate the local contribution from stray capacitances at bandwidths large enough for scanning. The setup allows experimenting between room temperature and $1.6\,$K and in high magnetic fields. We demonstrate the performance of the DSCM on gold pads buried under a flat layer of PMMA and by imaging the compressible strips that form at high magnetic fields and low temperatures at the edge of a Hall bar defined in a 2DEG.

\section{ACKNOWLEDGEMENTS}
We thank Prof M. Henini for growing the sample used in the quantum Hall experiment. We also acknowledge the valuable help of Dr F. Callaghan, Dr X. Yu and Prof A.J. Kent in earlier stages of this project. This work is financially supported by the Engineering and Physical Sciences Research Council (UK).

\bibliographystyle{apsrev}

\begin{thebibliography}{99}
\bibitem{footnote1}
\bibinfo{note}{This model can also be applied for three-dimensional conducting layers that are thin compared to the tip-radius.}
\bibitem{footnote2}
\bibinfo{note}{We neglect possible depletion due to not properly compensated work function differences or the tip voltage.}
\end{thebibliography}

\end{document}